\documentstyle{article}

\input{epsf}

\begin{document}
\hoffset=-1.5cm
\bibliographystyle{unsrt}

\centerline{\Large \bf The topological structure of 2D disordered cellular systems} 
\bigskip  
\centerline{{H. M. Ohlenbusch, T. Aste, B. Dubertret and N. Rivier} }
\smallskip
\centerline{\it Equipe de Physique Statistique, LDFC  Universit\'e Louis Pasteur,}
\centerline{\it  3, rue de l'Univresit\'e, 67084 Strasbourg France. }
\centerline{\it  helgo,benoit@maxent.u-strasbg.fr,  tomaso,nick@ldfc.u-strasbg.fr}

\date{\today}
\bigskip

\begin{abstract} 
We analyze the structure of two dimensional disordered cellular systems generated by
extensive computer simulations.  
These cellular structures are studied as topological
trees rooted on a central cell or as closed shells arranged concentrically around a germ
cell.  
We single out the most significant parameters that characterize statistically the organization of these patterns.
Universality and specificity in disordered cellular structures are discussed.
\end{abstract}



\section{Introduction}

Cellular structures, space-filling disordered patterns, are widespread in nature \cite{dAT,WR84,Stavans:0}.  
In an ordered structure, one must constrain the elementary bricks (or group of them) to satisfy
the local rotational symmetry compatible with translational invariance.  By contrast, a
disordered structure is free of any local symmetry constraint, and only subjected to the
inescapable, topological condition of partitioning space.  In two dimensions, these
structures (froths) are partitions of the plane by irregular polygons.  Disorder, or
absence of symmetry, imposes minimal incidence numbers (3 edges incident
on a vertex, 2 faces incident on a edge).  The Four-Corner boundary between the States of
Arizona, Utah, Colorado and New Mexico is a decision made in Washington; it has
nothing to do with population dynamics or with agricultural efficiency.  It is neither
topologically, nor structurally stable:  a small deformation splits it into two
topologically stable three-State vertices, found everywhere else on Earth.

\begin{figure}[htb]
\vspace{-1.cm}
\epsfxsize=8.cm
\hspace{1.6cm}
\epsffile{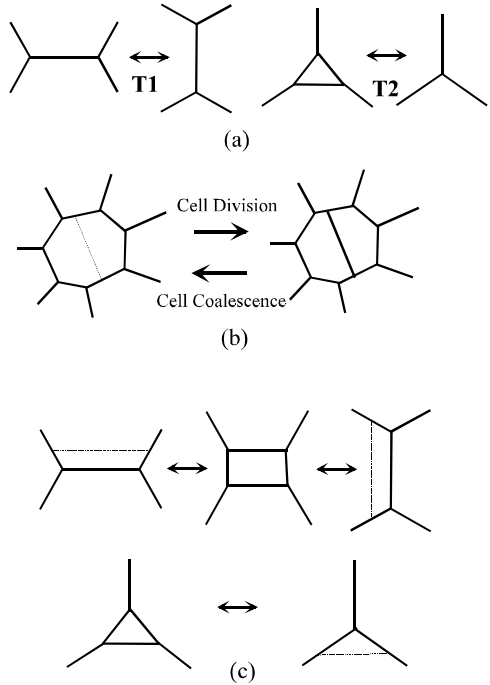}
\begin{centering}
\vspace{-1.cm}
\caption{ 
A froth is a cellular network generated or transformed by successive applications of two local 
elementary transformations: T1 (neighbours exchange), and T2 (triangular) cell disappearance {\bf (a)}.
These two transformations can also be regarded as combinations {\bf (c)} of cell division and 
cell coalescence {\bf (b)}.
 \label{f.TopTra}}
\end{centering}
\end{figure}

The set of all the possible configurations of a froth can be explored by elementary, local
topological transformations \cite{Alexander} which, in two dimensions, are:  {\bf T1}, exchange of 
neighbours (for example, a flip between the stable alternatives of UT-NM or of AZ-CO as neighbours, about
the unstable four-corner boundary); {\bf T2}, disappearance (or apparition) of a
triangular cell into a single vertex. 
Two alternative, equivalent transformations are the cell-division (called mitosis in biology) 
and cell-coalescence. 
It can be easily seen (fig.\ref{f.TopTra}) that local, specific combinations of division and coalescence 
generate the two elementary transformation T1 and T2, and vice versa.  
These elementary, local topological transformations are observed in all natural foams, soap 
froths, metallurgical grain aggregates, but also in biological tissues (epidermis) \cite{DR} 
or in the cellular networks produced by B\'enard-Marangoni convection, whether in the laboratory 
or on the surface of the sun (solar granulation) \cite{WR84,Stavans:0}.

In  infinite two dimensional (2D) Euclidean froths (or froths with periodic boundary conditions), the average number of 
neighbours per cell is fixed by the Euler relation to be equal to 6 \cite{WR84}.
Given a cell with $n$ edges, the quantity $q=6-n$ is called its topological charge.
The sum of the charges of the cells over the  whole infinite Euclidean froth must be equal to zero.
The topological charge cannot be generated or destroyed. 
It is just shuffled in between cells by the topological transformations.

\begin{figure}
\vspace{-8.cm}
\epsfxsize=16.cm
\epsffile{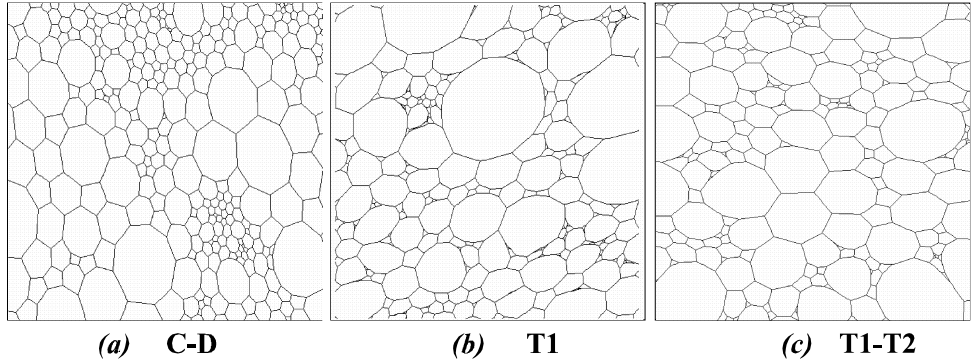}
\vspace{-9.cm}
\begin{centering}
\caption{Typical topological froths generated by: {\bf (a)} Cell division and coalescence; 
{\bf (b)} T1 transformations only; 
{\bf (c)} T1 and T2 transformations.   
\label{f.froths}\label{f.CDfroth}\label{f.T1froth}\label{f.T1T2froth}}
\end{centering}
\end{figure}

In this paper, extensive computer simulation are performed to obtain disordered
systems with more than $10^5$ cells generated from an hexagonal lattice and
applying at random either only T1, or T1 and T2, or cell-division and coalescence 
transformations.
In these simulated froths, we forbid configurations with cells which are neighbours of
themselves and cells with two edges.
The typical resulting structures are shown in fig.\ref{f.froths}.

The structural organization of these computer-generated froths are studied in terms of
the relative distances between cells.  
Two different topological distances can be introduced:  the {\it bond distance} \cite{bd98,dst97} 
and the {\it shell distance} \cite{RivPc,LeCearPc,Fortes,AsBoRi} (section \ref{TopDis}).  
These two distances are different: the bond distance is defined on the 
graph of the froth; the shell distance is best defined on its dual triangulation.
They are associated with the number of edges (or vertices) of the shortest paths
between two cells.
The relation between the two distances is very complex and strongly dependent on the froth 
organization, because they involve minimal paths.

\begin{figure}[htb]
\vspace{-2.cm}
\epsfxsize=8.cm
\hspace{3.cm}
\epsffile{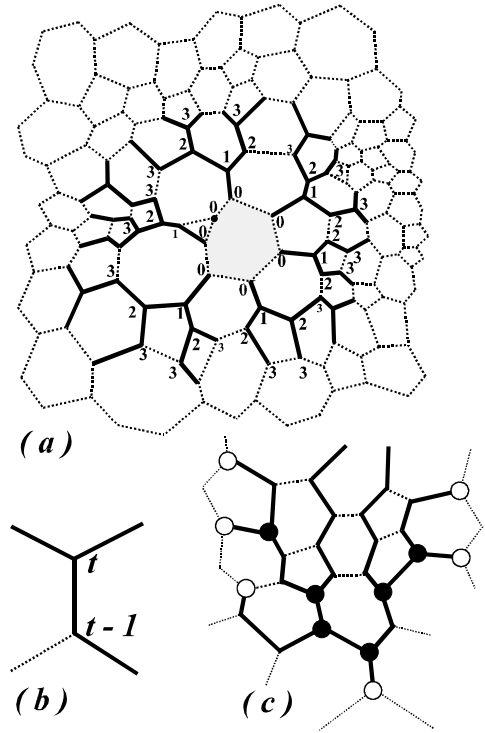}
\vspace{-1.cm}
\caption{ {\bf (a)} The forest of n bond trees. {\bf (b)} In a tree, any vertex at distance $t$ 
is connected with one bond to a vertex at distance $t-1$. 
Two of the three cells incident on a vertex at distance $t$ are also incident on a vertex
at distance $t-1$.
The mean vertex connectivity ($z_t$) tends asymptotically to 2, which is typical of a tree 
structure spanning vertices distributed uniformly (see also fig.\ref{f.hexaTree}).
In {\bf (c)} vertices with connectivity $z=1$ and $z=3$ are labelled with white and black 
circles respectivelly.
\label{f.bondTree}}
\end{figure}

The characterization of a disordered structure is a difficult task.  The absence of
translational order or repetitive local configuration makes in principle necessary to
know the position, size and shape of every cell in order to
completely characterize the structure.  
But, disorder makes most of this detailed local information unimportant.
The physical properties of the froth must be describable in terms of macroscopic 
statistical  information.
A simple and powerful way to study disordered cellular systems, is to analyze the
structure as organized around a given arbitrary central cell.  
The information about the structure can be obtained  in terms of the properties of the
set of cells at the a given distance (bound or shell) from the central cell.  
We show that the classification by bond distances reduces the froth to a set of trees 
rooted on the vertices of an arbitrary central cell (see fig.\ref{f.bondTree}).  
On the other hand, in terms of shell distance, the froth can be seen has a set
of closed layers of cells arranged concentrically around the central cell (see
fig.\ref{f.shell}).  
This classification in terms of distances naturally introduces a structure and a hierarchy in the froth.  
Moreover, a physical meaning is directly associated to the sets of equally-distant cells:  
any perturbation, signal or information propagates from a given cell to the whole froth through 
these structures and cells at the same distance are those reached -on average- at 
the same time and with equal intensity by the signal.
They are the successive wave fronts of the signal in the froth.

\section{Topological structure of a froth} \label{Stru}

\subsection*{Topological distances} \label{TopDis} 

Consider two cells in a froth.  These cells are connected by many paths along the
edges of the direct graph or of its dual triangulation.  
Two different topological distances can be associated with minimal paths:

{\bf The bond distance}, is the minimum number of 
edges necessary to connect two cells in the direct graph of the froth.\footnote{This path 
is a set adjacent edges ({\it `bonds' }) connecting the two cells.}.
(The concepts of bond distance and of bond neighbours was first introduced
by one of the authors \cite{bd98}, who called them T1 distance and T1 neighbours, 
respectively \cite{dst97}.)

{\bf The shell distance} (or simply, topological distance \cite{AsBoRi}),  is the minimum number of
edges connecting the centers of two cells in the dual triangulation.\footnote{The set of 
equidistant cells makes a closed layer around the central cell, and
successive layers form concentric rings.  The edges separating two successive layers form a closed
loop: the {\it `shell' } \cite{AsBoRi}.}

\noindent
All the cells in the froth can be classified in terms of distances (bond
\& shell) with respect to a given ``central'' cell.  

We denote by $F_t$ the number of cells which are at a bond distance $t$
 and by $K_j$ the number of cells which are at a shell distance $j$ from a given central cell.
These quantities are in general dependent on the number of edges $n$ of the central cell.
For any froth on the Euclidean plane, $F_t$ and $K_j$ increase with the
respective distances.  
But the rates of growth are characteristics of a specific cellular
system and are good, sensitive quantities to characterize its disorder 
(see section \ref{ToCar} and \cite{AST,ShapeM:1}).

\begin{figure}
\vspace{-4.cm}
\epsfxsize=10.cm
\hspace{1.5cm}
\epsffile{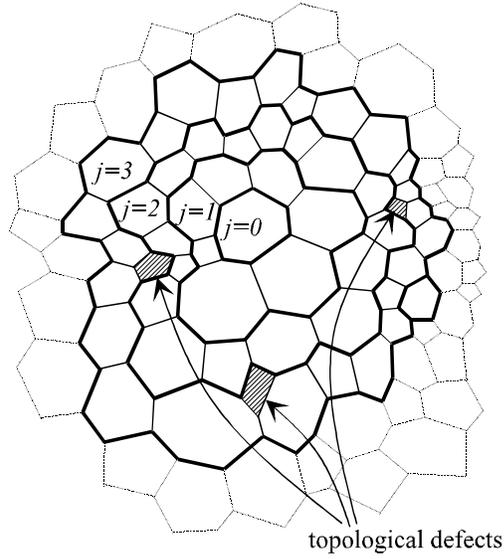}
\begin{centering}
\vspace{-3.cm}
\caption{The shell structure. 
Any froth can be analyzed as structured in a system of closed concentric layers of cells at the same
shell distance with respect to the central cell.  The cells (hatched in the figure) in layer $(j)$ 
with neighbours in layers $(j-1)$ and $(j)$ only, are local topological defects included between the layers.  
\label{f.shell}}
\end{centering}
\end{figure}

\subsection*{The n rooted bond trees} \label{VtFt}

Associated with the bond distance, there is a spanning forest of n trees
rooted on the central cell and reaching every vertex of the froth in the 
least number of steps .
These trees are the minimal bond paths which join the vertices of the
froth with the central cell.
Consider a vertex of the froth at a large bond distance $t$.
It has 3 neighbouring vertices, with at least one at a distance $t-1$.  
Select one of them and connect it with a {\it bond}. 
The other neighbouring vertex at  $t-1$ (if there is one) is left unconnected. 
Continue down in distances until a vertex of the central cell is reached. 
This is the {\it root} of the tree, which, at this stage, is simply a branch (chain) of $t$ bonds. 
Then, begin again with another vertex at distance $t$, down in distances. 
This second $t$-branch either meets the first at some point, or it reaches another root (vertex 
of the central cell). 
Proceed successively with every vertex at distance $t$, down in distances until the branch 
connects another  branch or is rooted into the central cell. 
Consider then the vertices at distance $t-1$ which have not yet been reached by the
branches already constructed, and repeat the operations above. 
By repeating the process to all the vertices yet unspanned at distances $t$,
$t-1$, $t-2$,....,$2$, $1$, $0$, we generate a forest of $n$ trees, where
$n$ is the number of vertices (and of edges) of the central cell. 
These trees are rooted on the central cell and span all the vertices of the
froth. 
Fig.\ref{f.bondTree} (a) shows a forest of trees up to distance 3.
Each vertex in the tree has connectivity $2 \le z \le 3$, except the root
and branch ends which have connectivity $1$ (zero if the tree consists of
its root vertex only, as happens, for example, if the central cell has a
3-cell as topological neighbour). 
Call $z_t$ the average vertex connectivity at distance $t$ ($1 \le z_t \le
3$ for $t>0$ and $0 \le z_0 \le 1$).
It follows immediatelly that $V_{t+1} = (z_t - 1) V_t$ and $V_1 = z_0 V_0 =z_0 n $.
The number of vertices at distance $t > 1$ is therefore related to the number of
edges of the central cell ($n$)
\begin{equation}
    V_t = \Big[ z_0 \prod_{i = 1}^{t-1} (z_i - 1) \Big] n \;\;\;\;\; .
\label{Vt}
\end{equation}

The bond distance associated to a cell is  the bond distance of its vertex nearest to 
the central cell.
Consider a vertex at distance $t$. 
It has 3 incident cells. 
Two must also be incident on a vertex at distance $t-1$  
(they are separated by the bond linking vertex $t$ to vertex $t-1$, see fig.\ref{f.bondTree}(b)~).
Therefore, to each vertex at a given distance, there corresponds no  more than one cell at
the same distance, i.e. $F_t = \nu_t V_t$ with $\nu_t \le 1$.
In 2D Euclidean froths, a cell has 6 vertices on average, and 3 cells are
incident on any vertex.
Therefore we expect $\nu \simeq 0.5$ on average.

A special case is the hexagonal lattice where
there are 6 vertices and 6 cells at distance $t=1$, 12 vertices and 6 cells at
$t=2$, 12 vertices and 12 cells at $t=3$, and, in general $V_t = F_t = 3 (t + 1)$ for $t$ odd
and $V_t = 3 (t + 2)$ and $F_t = 6$ for $t$ even.
We have therefore $\nu_t = 0.5$ on average, with $\nu_t=1$ when $t$ is odd and $\nu_t=0$
when $t$ is even.
This difference between even and odd distances is characteristic of the
hexagonal lattice.
When the froth is disordered, this difference between even and odd distances
disappears and, asymptotically, the quantities $V_t$ and $F_t$ grow linearly with the
distance $t$ with $\nu_t \simeq 0.5$.
The growth law depends on the specific system and it is studied in section \ref{Sim}
and  \ref{ToCar}.

\begin{figure}
\vspace{-5.cm}
\epsfxsize=12.cm
\hspace{.5cm}
\epsffile{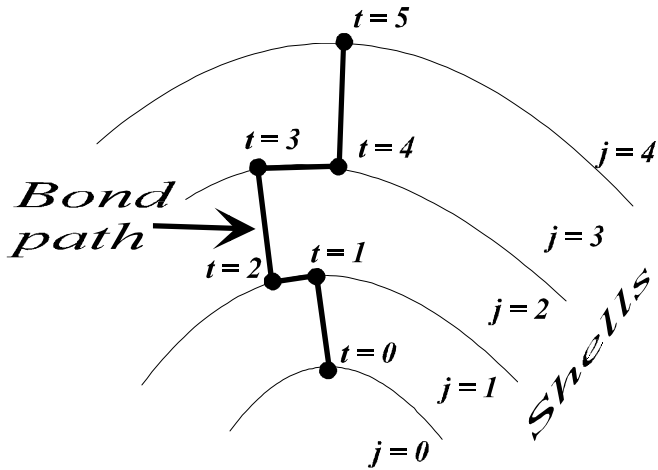}
\begin{centering}
\vspace{-7.cm}
\caption{The shortest bond path  crosses at least $t_{min} = 2 j -3$ vertices to connect the
central cell with a cell in the layer $j$.   \label{f.minBpath}}
\end{centering}
\end{figure}

\subsection*{The shell structure}

The structure associated with the shell distance for a typical froth is shown in fig.\ref{f.shell}. 
The shell distance foliates the froth into successive layers of cells
surrounding the central cell.
Most cells at a (shell) distance $j$ from the central cell have neighbours
at distances $j-1$, $j$ and $j+1$. They belong to layer $(j)$.
In addition, some cells at distance $j$ (hatched in Fig.\ref{f.shell}) have neighbours at distances 
$j-1$ and  $j$ only. 
These cells are local {\it defect } inclusions, intercalated between layers $(j-1)$ and $(j)$. 

In this paper we characterize the shell structure by the number of
cells at distance $j$,  $K_j(n)$, averaged over all $n$-sided central cells.
In the perfect honeycomb (hexagonal froth), $K_j = 6j$, there are no defects, and the successive 
layers have an hexagonal shape. 
In a random froth, the layers are roughly circular, but wiggly. 
At large distances,  $K_j (n)$ is larger than for the honeycomb, but still increases linearly with the
distance.
The rate of increase is related  to the degree of disorder and to the number of
topological defects in the system~\cite{AST}.

\subsection*{The combined structure}

The quantities $F_t$ and $K_j$ are both radial properties with the angular dependence
averaged out.  
In order to extract information on the angular fluctuations, it is interesting to study 
the interplay between bond and shell structures.
A bond tree develops radially outward from the central cell along edges which are
perpendicular to the shells and it broadens sideways along edges parallel to the shells.  
The distribution in the bond tree of edges parallel to the shells gives therefore a
measure of its spread.  
Call $N(j,t)$ the number of couples of cells  that are
simultaneously at a shell distance $j$ and at a bond distance $t$ from each other.  
(Clearly, $\sum_t N(j,t) = K_j N$ and $\sum_j N(j,t) = F_t N$, with $N$ the total number of cells in the
 system.)

To calculate $N(j,t)$, let one of the cells be the central cell. 
The other cell is at shell distance $j$. 
What is the bond distance $t$ between the two cells? 
It is readily seen (Fig.\ref{f.minBpath}) that at least 2 bonds are necessary to
go from one layer to the next. 
Thus, $t_{min} = 2j-3 = t_R$ for $j"\ge 2$. 
Therefore, $t_R$ is the radial component of the bond distance. 
(In an hexagonal froth, $t = t_R$, for $j"\ge 2$.)

In disordered froths, the bond distance $t$ is, in general, much larger
than $t_R$.
Suppose, for instance, that a path takes in average $\langle v \rangle$ extra bonds from one layer
to the next.
These extra bonds correspond to segments of the bond tree parallel to
the shell and also to extra vertices associated to the topological defects
between the shells.
From the central cell to a cell at distance $j$, there are $\langle v \rangle (j-1)$ additional bonds
and the total bond distance is therefore $t = t_{R} + \langle v \rangle (j-1) = 
(2+ \langle v \rangle) j -3-\langle v \rangle$.
The number $v$ of extra bonds between layers is a random variable with probability 
distribution $w(v)$ and mean $\langle v \rangle$.
When $w(v)$ is  Poissonian, the conditional probability $P(t|j)$ of finding a cell at a bond
distance $t$, given that it lies at a shell distance $j$, is a  shifted Poissonian with average 
$\bar t = \langle v \rangle (j-1) +2j-3$ and variance $\sigma^2 = \langle v \rangle(j-1)$ (see appendix \ref{ComDis}).
The number of couples of cells which are simultaneously at a shell distance $j$
and at a bond distance $t$ is therefore

\begin{equation}
N(j,t)=N K_j P(t|j) = N K_j  {e^{- \langle v \rangle (j-1)} [\langle v \rangle (j-1)]^{t-2j+3}
                                   \over (t-2j+3)! }\;\;\;\;\; .
\label{N(j,t)}
\end{equation}

The arrangement of the cells in any layer can therefore be described in terms of a single parameter 
$\langle v \rangle$ only.

\section{Simulation of disordered cellular systems } \label{Sim}

\subsection*{Coalescence and Division (C-D simulation)}

The  disordered cellular system is generated staring form an hexagonal lattice of 
$N_0= 144400 $ cells, with periodic boundary conditions, and performing $5 N_0$ of T1 transformations 
on edges chosen at random.  
Then, a cell chosen at random, is divided in two parts (mitosis) if it has more
than 6 sides or fused by coalescence with one of its neighbours  if has less than 6 sides.  
If it has six sides, it is divided with probability $p_d = 1/4$ or left unchanged with probability 
$1-p_d = 3/4$.  
The coalescence is performed between the cell and one of its neighbours chosen at random.  
We chose a symmetric mode of division:  the two ``daughters'' of an $n$-sided
``mother'' cell have both $2 + n/2$ edges for $n$ even, and  $2 + (n + 1)/2$
and  $2 + (n - 1)/2$ edges if $n$ is odd (which one has which is assigned
at random).

\begin{figure}
\vspace{-5.cm}
\epsfxsize=12.cm
\hspace{1.cm}
\epsffile{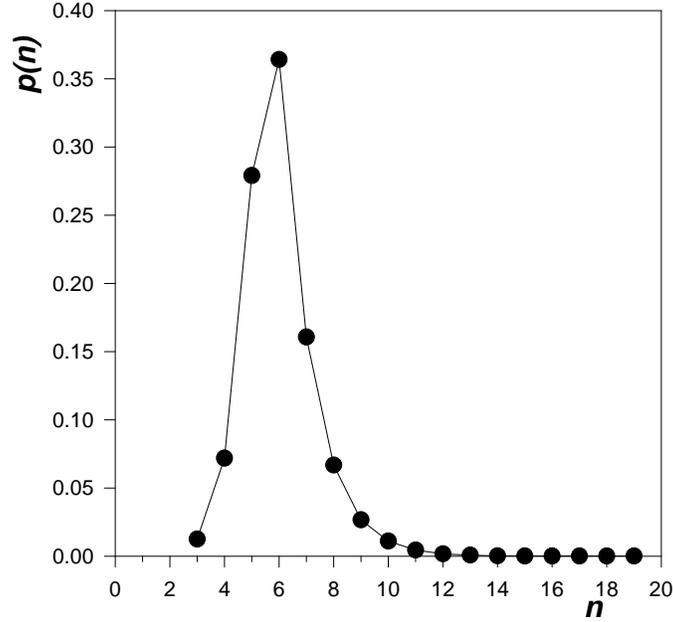}
\begin{centering}
\vspace{-4.cm}
\caption{Distribution of the number of edges per cell in the froth generated by applying  
cell-division and coalescence transformations (C-D simulation). \label{f.CDpn}}
\end{centering}
\end{figure}

After iterating the coalescence-division operations for $10 N_0$ times, the resulting structure 
(shown in fig.\ref{f.froths}(a)) reaches a stationary number of $N= 114475$ cells.
The normalized distribution of the number of edges per cell ($p(n)$) is shown in fig.\ref{f.CDpn}.
It has a maximum at $ n =6 $ and decreases exponentially for  $n \ge 6$) as
$p(n) = p_6 \exp[-\alpha (n-6)]$ with $p_6 = 0.37$ and $\alpha =0.9$.
The second moment of the distribution is $\mu_2 = 1.77$, and the Aboav-Weaire parameter 
 \cite{WR84,Aboav,Wea} is $a = 0.64$.

\begin{figure}
\vspace{-5.cm}
\epsfxsize=12.cm
\hspace{1.cm}
\epsffile{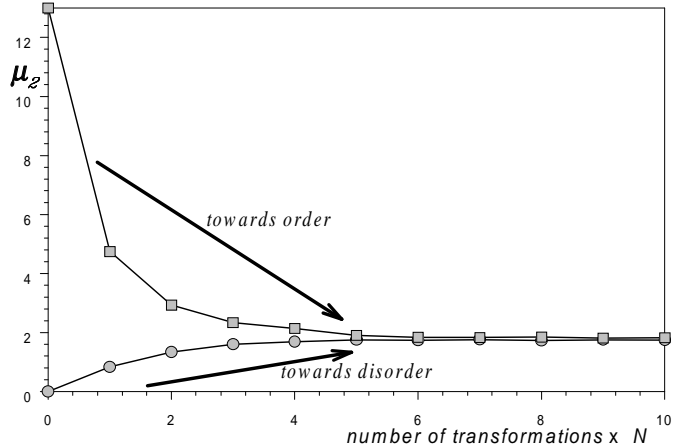}
\begin{centering}
\vspace{-5.cm}
\caption{Second moment of the edges distribution ($\mu_2$) v.s. the number of elementary topological 
transformations, in the system generated by cell-division and coalescence transformations (C-D simulation). 
The system converges toward a steady state which is independent of the starting configuration.
If the simulation starts from a very disordered state, the system self organizes reducing the disorder
by random local transformations. 
\label{f.CDmu2}}
\end{centering}
\end{figure}

This distribution is stationary under further division-coalescence transformations. 
It is also independent of the initial configuration. 
We have obtained statistically identical froths starting from the hexagonal lattice ($\mu_2=0$) or from a very
disordered froth with $\mu_2 >  10$. 
In the first case, the system evolves from  order to disorder during the simulation. 
By contrast, in the second case, the disorder of the system decreases throughout the simulation (see fig.\ref{f.CDmu2}).
This is an example of self organization induced by local random transformations only.
The distribution $p(n)$ reaches a steady state, regardless of the
parameter $p_d$ which controls the fate of 6-sided cells. 
However, the number of cells is very sensitive to $p_d$. 
For $p_d = 1/4$, once the steady state is reached, the number of cells in the froth remains also stable. 
But, when $p_d = 0 $, the  number of cells in the system decreases, whereas, when $p_d = 1/2$, it increases.

\subsection*{Random T1 (T1 simulation)}

An hexagonal lattice of $N = 100172 $ cells and periodic boundary
conditions is disordered by flipping (T1 transformation)  $10 N $ edges chosen at random.
(Note that the T1 transformation conserves the total number of cells in the
system.)  
If the chosen edge is bounding a triangular cell, the T1 transformation is refused
(lest it would generate a cell with two sides). 
The probability for a T1 to be refused increases during the
disordering process, it reaches the value of 0.21 at the end of the simulation.

\begin{figure}
\vspace{-5.cm}
\epsfxsize=12.cm
\hspace{1.cm}
\epsffile{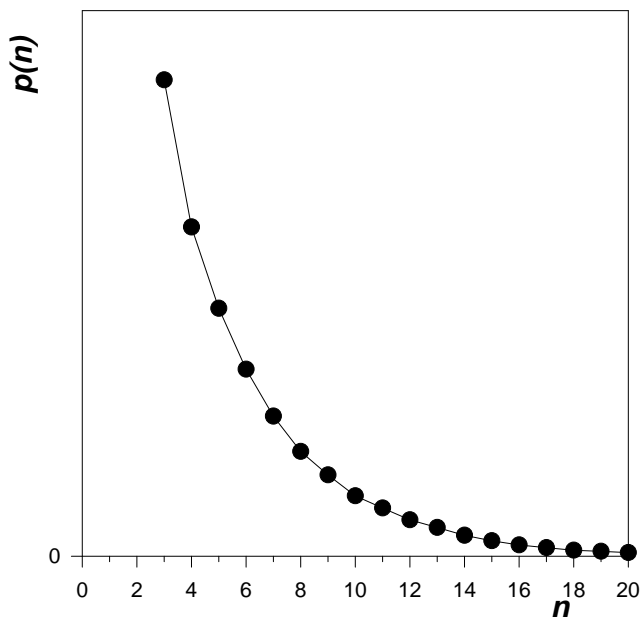}
\begin{centering}
\vspace{-4.cm}
\caption{Distribution of the number of edges per cell in the froth generated by T1 transformations 
(T1 simulation). \label{f.T1pn}}
\end{centering}
\end{figure}

A typical cellular system obtained with this simulation is shown  in fig.(\ref{f.T1froth}b).
Fig.\ref{f.T1pn} shows the probability distribution $p(n)$ for this froth. 
The second moment is $\mu_2 = 11.9 $ and the Aboav-Weaire parameter is $a = -0.79 $.
The average number of sides per cell is equal to 6 (as imposed by the Euler theorem \cite{WR84}) but the 
distribution is an exponential with a maximum at  $n=3$,
$p(n) = p_3 \exp[-\alpha (n-3)]$ with $p_3 = 0.27$ and $\alpha =0.30$.
In this simulation, most of the cells are triangles because it is easier 
to generate a triangular cell than  to make it gain an edge by random T1.

\subsection*{Random T1 and T2 (T1-T2 simulation)}

\begin{figure}
\vspace{-1.cm}
\epsfxsize=12.cm
\hspace{1.cm}
\epsffile{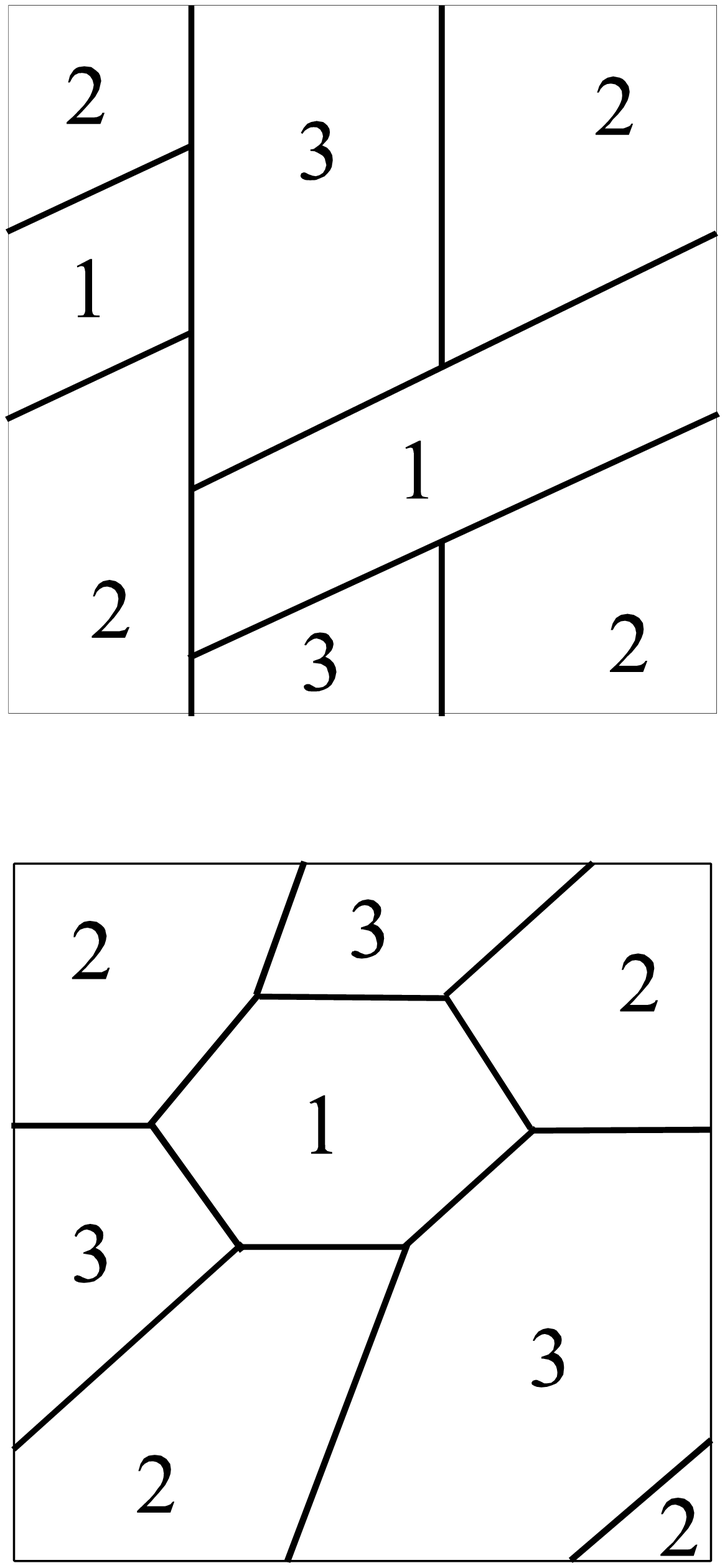}
\begin{centering}
\vspace{-1.cm}
\caption{   Two topologically equivalent views of the  final configuration for the T1-T2 simulation. 
There are three cells of 6 sides tiling a torus. 
No further T1 or T2 transformations can be applied on this final, minimal configuration without
generating a cell that is neighbour  of itself.  \label{f.min_tour}}
\end{centering}
\end{figure}

In this type of simulation,  a flip (T1 transformation) is performed
on any edge chosen at random, except if the edge is bounding a
triangular cell. In this case, the triangular cell is made to
disappear (T2 transformation). For this reason, the number of cells 
decreases during the simulation.  
We found that, when $5 N_0$ transformations are
applied on an initially hexagonal lattice of $N_0$ cells, the number of cells is reduced 
by half. 
But $10 N_0$ transformations are more than enough to reduce the froth to 3 (hexagonal) cells only.
This is the minimum possible number of cells for a froth tiling a torus.
Here each cell has 2 neighbours and 6 edges and they are arranged as shown in fig.\ref{f.min_tour}.
Note that, for this configuration, there are two families of closed paths of only 4 edges, not bounding a cell,
that wind around the torus.
No further T1 or T2 transformations can be applied on this final, minimal configuration without
generating a cell that is neighbour  of itself.

\begin{figure}
\vspace{-5.cm}
\epsfxsize=12.cm
\hspace{1.cm}
\epsffile{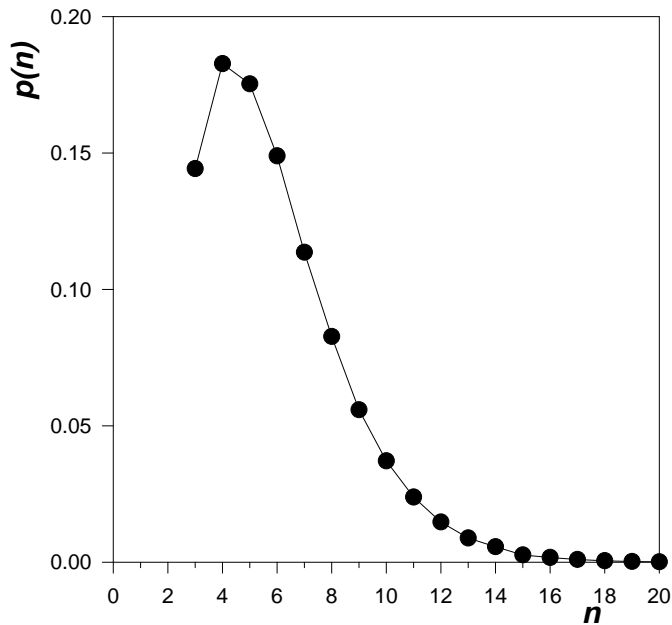}
\begin{centering}
\vspace{-4.cm}
\caption{Distribution of the number of edges per cell in the froth generated by T1 and T2 transformations 
(T1-T2 simulation). \label{f.T1T2pn}}
\end{centering}
\end{figure}

This simulation started with an hexagonal lattice of $N_0=250000$ cells
on which we applied $5N_0$ of T1 or T2 transformations on edges chosen at random.
The resulting structure had a number of cells $N = 115149$.
A typical froth generated by T1-T2 transformations is shown in fig.\ref{f.T1T2froth}(c).
The probability distribution $p(n)$ is plotted in fig.\ref{f.T1T2pn}.  
It is quasi-stationary (until the system is close to its final configuration
of 6 hexagonal cell).
It has a maximum in $n=4$ and then decreases with  an exponential tail  
$p(n) = p_4 \exp[-\alpha (n-4)]$ with $p_4 = 0.84$ and $\alpha =0.53$, for $n > 7$. 
The second moment is $\mu_2 = 6.33 $ and the Aboav-Weaire parameter is $a = 0.05$.

\section{Results and Discussion  }\label{ToCar}

For each simulation, the cellular systems are characterized by the number of cells in the shell structure ($K_j$), 
in the bond trees ($F_t$) and in the combined structure ($N(j,t)$), at
the respective distances $j$ and $t$ of the central cell. 
The number of vertices in the bond trees $V_t$ and their mean coordination $z_t$ are also investigated.

\begin{figure}
\vspace{-4.cm}
\epsfxsize=12.cm
\hspace{1.cm}
\epsffile{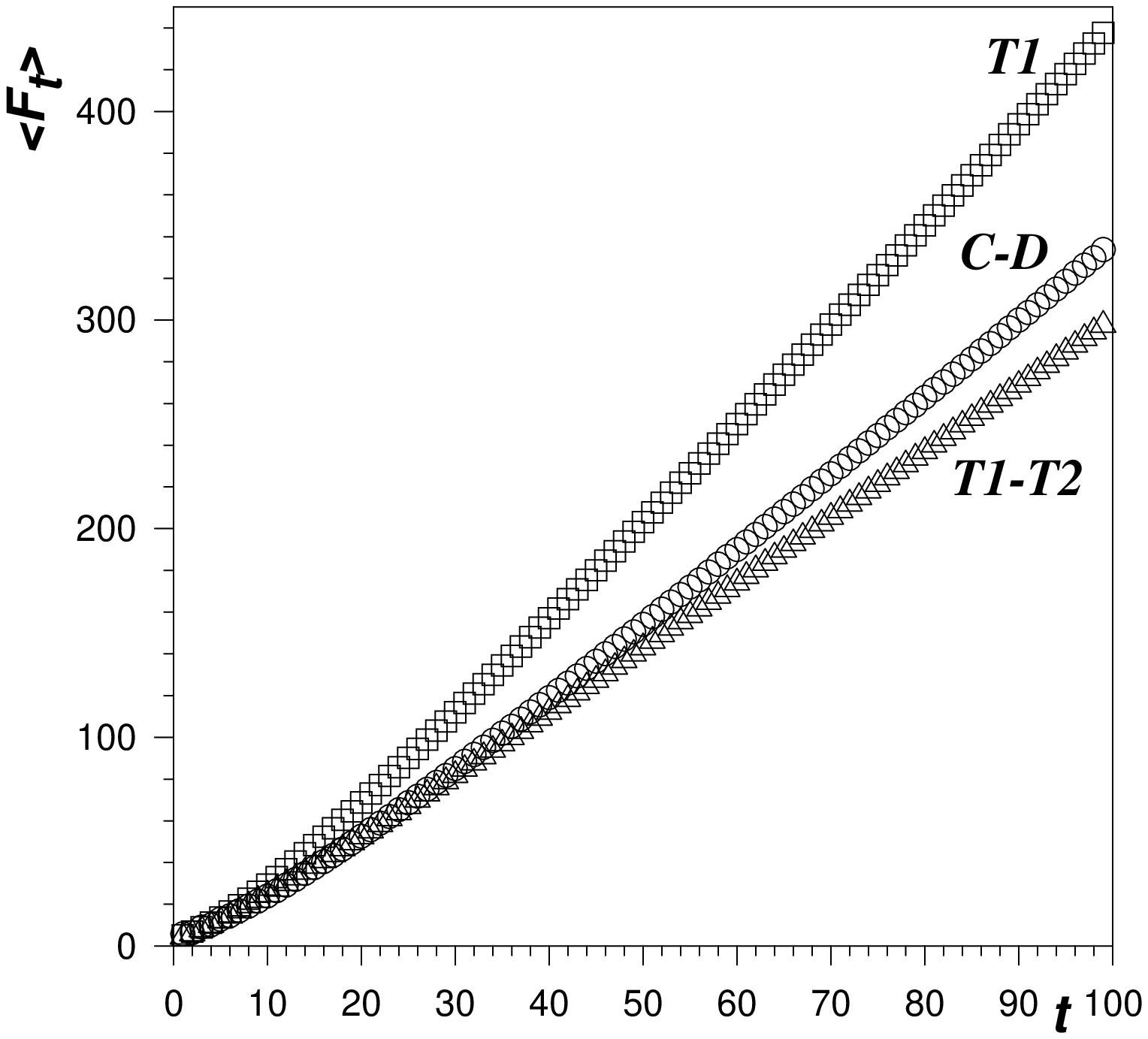}
\begin{centering}
\vspace{-4.cm}
\caption{Average number of cells $\langle F_t \rangle$ at distance $t$  in the bond trees, for the three 
simulations C-D, T1 and T1-T2. The average is over all central cells. \label{f.Ft}}
\end{centering}
\end{figure}

\subsection*{Bond trees}

The averaged number of cells in the bond tree ($\langle F_t\rangle =\sum_n p(n) F_t$) 
as a function of the bond distance $t$ is shown in fig.\ref{f.Ft}.
For the three sets of simulations, $\langle F_t\rangle$ increases linearly with the distance 
as $t \rightarrow \infty$, with a slope $3.577$ (D-C), 4.68 (T1) and 3.112 (T1-T2).
As anticipated in Section  \ref{VtFt}, the number of vertices at distance $t$ in the bond tree 
is expected to be proportional to the number of cells at that distance, $V_t = F_t/\nu_t$.  
Indeed, we find a coefficient of proportionality that asymptotically tends to $\nu_t \rightarrow \nu_\infty = 0.5$.

\begin{figure}
\vspace{-6.cm}
\epsfxsize=11.cm
\hspace{1.cm}
\epsffile{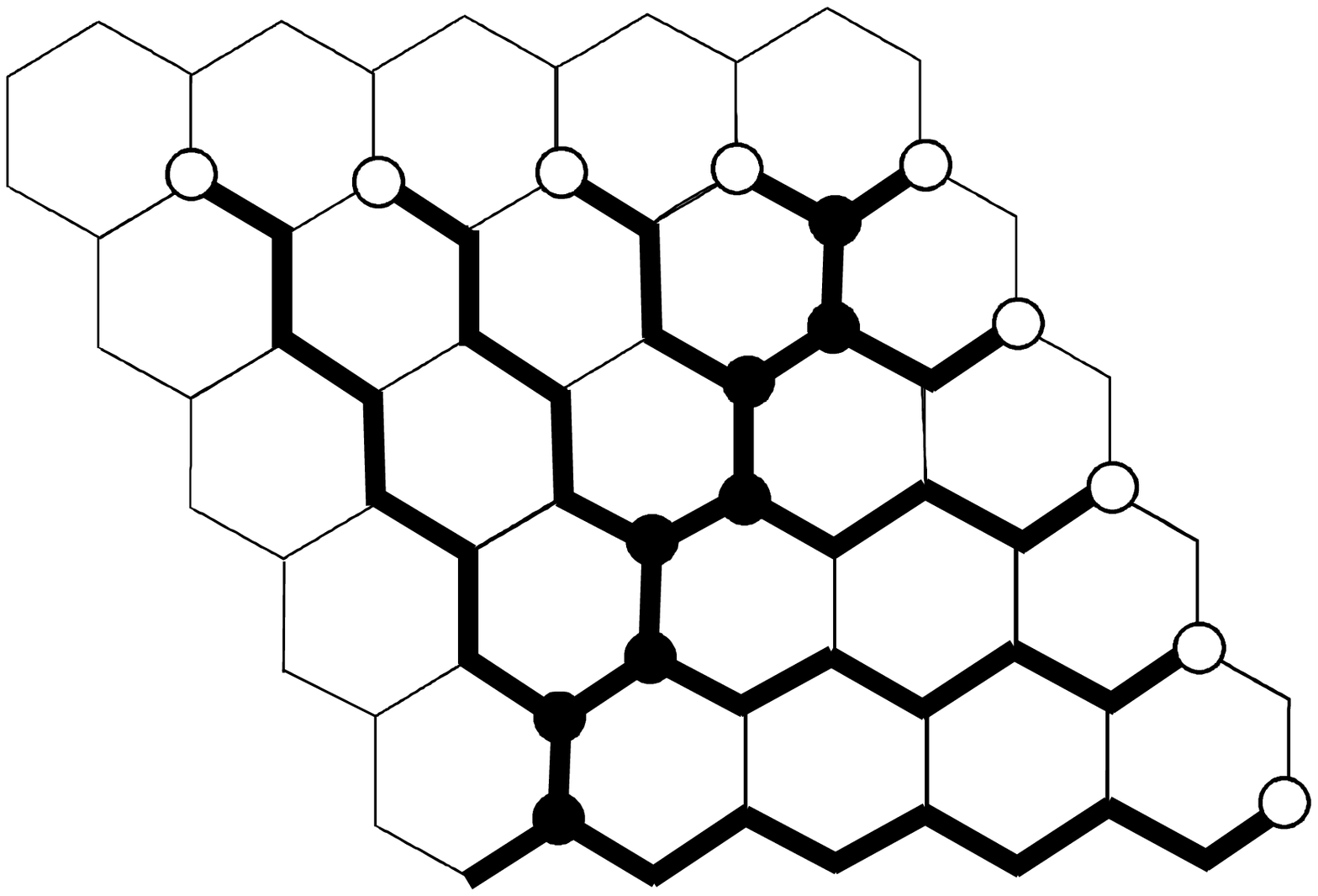}
\begin{centering}
\vspace{-4.cm}
\caption{A bond tree spanning $1+t(t+1)/2$, uniformly distributed vertices
(here on an hexagonal tiling). There are $t$ vertices at distance $t$, at
the end of the branches, with coordination 1 (white circles); the t vertices on the
main trunk have coordination 3 (black circles); the root has coordination 1. The
remaining $t(t-3)/2$ vertices have coordination 2. Thus $\langle z \rangle = 2 - O(1/t^2)$.
\label{f.hexaTree}}
\end{centering}
\end{figure}

\begin{figure}
\vspace{-6.cm}
\epsfxsize=11.cm
\hspace{1.cm}
\epsffile{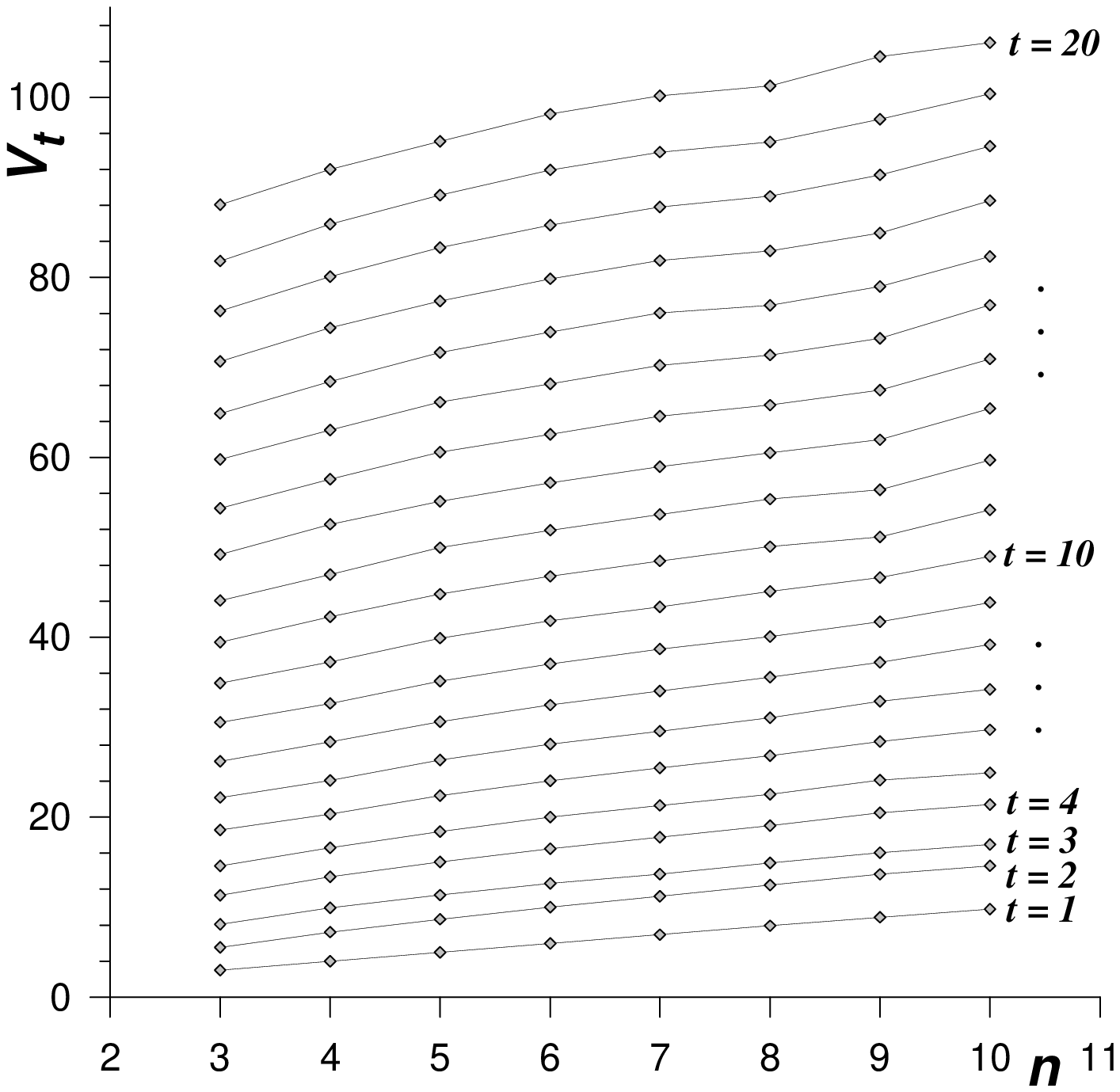}
\begin{centering}
\vspace{-4.cm}
\caption{Number of vertices $V_t$ in the bond trees as function of the number 
of edges of the central cell $n$, for several bond distances $t$.
The data refer to the C-D simulation.
Similar behaviours were found for the other two types of simulations. 
\label{f.VtCD410}}
\end{centering}
\end{figure}

Asymptotically, the mean vertex connectivity $z_t$ tends to 2, which is typical
of a tree structure spanning vertices distributed uniformly on a finite
area (see fig.\ref{f.bondTree}(c) and fig.\ref{f.hexaTree}).
There are $n$ such trees for an $n$-sided central cell. 
If the trees were not impeding each other one would expect $V_t \propto t n$, asymptotically in $t$. 
But trees do impede each other. 
This interaction between the $n$ trees implies that, after a certain distance,
the leading contribution to $V_t$ is proportional to $t$, regardless  of $n$.
The influence of the central cell is screened by the disorder \cite{RiDuAsHo}, which
distributes the vertices uniformly on the plane. 
This is manifest in Fig.\ref{f.VtCD410}, which shows that $V_t$ is linear in $n$, 
with a slope independent of $t$.

\subsection*{Shell structures}

\begin{figure}
\vspace{-3.cm}
\epsfxsize=10.cm
\hspace{1.cm}
\epsffile{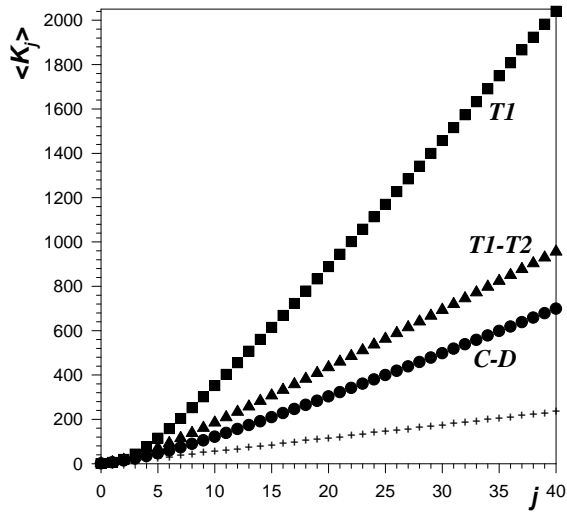}
\begin{centering}
\vspace{-3.cm}
\caption{Average number of cells $\langle K_j \rangle$ at distance $j$ in the shell layers, for 
	the three simulations C-D, T1 and T1-T2 and for the hexagonal lattice ($+$ symbol).
	The average is over all central cells. 
	\label{f.Kj}}
\end{centering}
\end{figure}

The number of cells $K_j$ at shell distance $j$ (whether within
layer $(j)$ or inside defect inclusions) increases linearly
with $j$ after the first few layers.
The dependence of $K_j$ on the number of edges $n$ of the central cell is also linear.
Among all cells at distance $j$, the proportion belonging to defects is 18\% for the
C-D simulation, 52\% and 38\% for the T1 and T1-T2 simulations,
stationary after a few layers. 
Fig.\ref{f.Kj} shows the averaged (over all central cells) number of cells
$\langle K_j \rangle$ at a shell distance $j$.
Figure \ref{f.Kj} shows clearly that the rate of  growth $C$ of 
$\langle K_j \rangle$ is very different in the three types of simulations.
The number of cells at a given shell distance is larger in the more
disordered systems. 
Specifically, the rate is $C = 19.35$ for the C-D simulation ($\mu_2= 1.77$), $C= 56.6$ for the T1
simulation ($\mu_2= 11.9$) and $C = 25.7$ for the T1-T2 simulation ($\mu_2= 6.33$). 
These values are much larger than $C = 6$ for the hexagonal tiling ($\mu_2= 0$), or than $C = 2 \pi $
expected if layers were circular annuli.
Indeed, in disordered froths the concentric layers wiggle around the averaged circular annulus. 
This behaviour has already been observed in soap and Voronoi froths \cite{AST}.
It can be interpreted as an additional, negative curvature caused by the disorder, and compensated 
by the positive curvature of the defects intercalated between the layers to produce a tiling which is
globally Euclidean. 
A relation for the slope $C$ in  froths free of topological defects (SSI) ans with shortest 
ranged correlations was obtained in \cite{AST}: $C = 6 + (3-a)\mu_2$, with $a$ the Aboav-Weaire parameter.
This relation give $C= 10.2$, $C = 51.3$ and $C = 24.7$ for the C-D, T1 and T1-T2 simulations, respectivelly.
The moderate agreement indicates that defects are relevant.

\begin{figure}
\vspace{-3.cm}
\epsfxsize=10.cm
\hspace{1.cm}
\epsffile{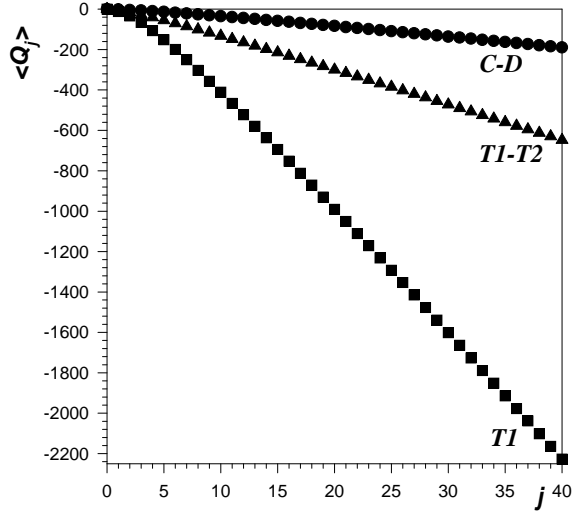}
\begin{centering}
\vspace{-3.cm}
\caption{Average topological charge $\langle Q_j \rangle$ v.s. $j$, 
within a cluster bounded by a layer of cells at shell distance $j$.
The average is over all central cells. 
The topological charge of a $n$-sided cell is $6-n$   \label{f.Qj}}
\end{centering}
\end{figure}

The topological charge ($Q_j$) of a cluster bounded by layer $(j)$ (included) has been
measured in the three types of simulation.
Figure (\ref{f.Qj}) shows the averaged charge $\langle Q_j \rangle$ as a function of the shell distance $j$ 
(the average is over all the central cells). 
The topological charge  $\langle Q_j \rangle$ is negative and decreases linearly with $j$ (similar results were
found for soap and Voronoi froths \cite{AST}).
The mean topological charge per defect ($\langle q^{def} \rangle$) is independent of $j$ at large $j$.
We found $\langle q^{def} \rangle = 1.1$ in the C-D simulation, 
$\langle q^{def} \rangle = 2.1$ and $\langle q^{def} \rangle = 1.8$ in
the T1 and T1-T2  simulations, respectively.

The negative topological charge of the cluster ($Q_j$) is due to the wiggling of the boundary.
The linearity shows that the amplitude of the wiggling remains constant with increasing shell distance.
The boundary does not roughen as $j$ increases.
The negative charge of the cluster is balanced by the positive charges of the defects just outside its boundary.

\subsection*{Combined structure}

\begin{figure}
\vspace{-4.cm}
\epsfxsize=16.cm
\hspace{1.cm}
\epsffile{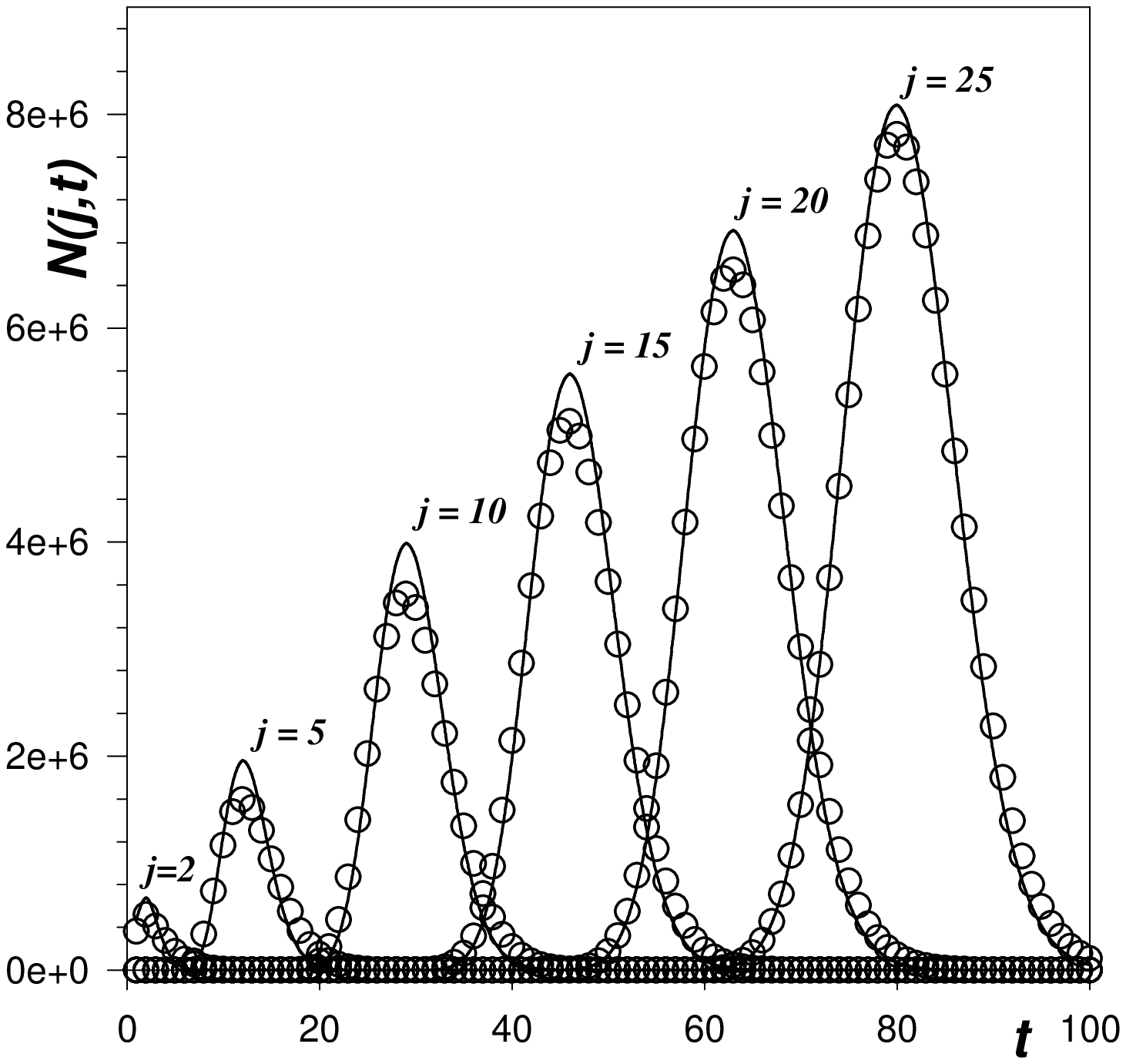}
\begin{centering}
\vspace{-4.cm}
\caption{Number of couples of cells $N(j,t)$ at a shell distance $j$ and bond distance $t$, as 
function of $t$ for several values of $j$.
The symbols are data taken from the T1 simulation.
The full lines are the theoretical solution Eq.(\ref{N(j,t)}) with $\langle v \rangle = 1.39$.
Note the linear correlation between the most probable value of $t$ and $j$.
\label{f.Njt}}
\end{centering}
\end{figure}

The distribution $N(j,t)$ of couples of cells at shell distance $j$
and bond distance $t$ is represented  in fig.\ref{f.Njt} as a function
of $t$ for different values of $j$, for the T1 simulations. 
Beyond $j = 10$, the distribution is symmetric, peaked at
the mean value $\bar t = 3.39  j - 4.2$, with a variance 
$\sigma^2 = 1.35 j + 3.2$.
The other two types of simulation exhibit similar behaviours. 
The theoretical distribution (a shifted Poissonian, eq.(\ref{N(j,t)}), derived in Appendix  \ref{ComDis}) is also 
plotted in fig.\ref{f.Njt} (full line), with only one parameter  $\langle v \rangle$ to be fitted 
($\langle v \rangle$ measures the mean increase in lateral spread of the tree, due to disorder, it is defined in Section \ref{Stru}.)
The agreement is excellent. 
We obtain $\langle v \rangle = 0.31$ (C-D simulation), $\langle v \rangle = 1.39$ (T1 simulation) and 
$\langle v \rangle = 0.84$ (T1-T2 simulation). 
As expected, branches are longer in the more disordered systems.

\section{Conclusions}
We have simulated and analyzed very different cellular patterns 
which have only the name of froth in common: they are planar networks
with minimal vertex connectivity.
Several disordered froths with more than $10^5$ cells were  generated
by computer simulations.
The results presented in this paper refer to three types of simulations of cellular
systems (C-D, T1 and T1-T2) generated with different techniques, which exhibit different degrees 
of disorder.
The T1 and T1-T2 simulations generate very disordered systems where the
number of edges of a cell fluctuates widely. (The sizes of the cells
have not been relaxed or adjusted in our topological simulations).
The C-D simulation generates a more  homogeneous cellular pattern, somewhat akin to soap
froths or other natural cellular structures. 
When C-D simulations are applied on an initially very disordered
froth, the system self-organises, and a more homogeneous configuration
with steady statistical properties is reached. 
In all three types of simulation, a stationary distribution of cells can be obtained. 
Moreover, the number of cells in the froth also remains stationary in the C-D
simulation for a special value of the probability $p_d = 1/4$ of dividing
6-sided cells.

These disordered cellular structures were characterized  in terms of two
distances from an arbitrary central cell: the {\it bond} and the {\it shell} distance.
In terms of bond distances, the froth is structured in a forest of $n$ {\it bond trees}, rooted 
on the $n$ vertices of the central cell.
They have asymptotic vertex coordination $z_\infty = 2$ and the number of vertices
$V_t$ at a given distance $t$ increases linearly both with $n$ and with $t$.

The number of cells $K_j$ at a shell distance $j$ from a given central cell,
increases linearly with  $j$.
This linearity seems to indicate that froths remain Euclidean
structures, and that the concentric layers of cells at the same cell
distance $j$ from the central cell, if they wiggle, do not roughen. 
There is no indication of fractal behaviour in our simulations (see \cite{Del-LeCaer} for an 
-artificial- example of fractal froth).
The rate of increase ($C$) of the number  $K_j$ of cells with the shell distance $j$ varies
strongly with the type of simulation.
It is larger for the more disordered systems, $C=6$ for the hexagonal tiling, and $C\simeq 19$, 25 
and 56 for the C-D, T1-T2, and   T1 simulations, respectively. 
The slope of $K_j$ v.s. $j$ is therefore a very good parameter to differentiate froths with
various degrees of disorder.
Moreover, these differences should be observable in the physical properties
of the froth like the diffusion coefficient \cite{AstRW97}.

The topological charge $Q_j$ of a cluster of cells at shell distances $i \le j$
from a given central cell is negative, with a linear dependence on  $j$.
This shows that layers of equidistant cells wiggle without roughening at
larger distances, that this negative topological charge or effective
curvature is entirely due to the outermost layer, and that it is 
completely balanced by the positive charges of the topological defects
just outside the outermost layer of the cluster.

In a given froth, the two topological structures {\it bond trees} and {\it
shell structure} are intimately related.
The combined structure was studied in terms of the number $N(j,t)$ of cells which
are simultaneously at a shell distance $j$ and at a bond distance $t$ from a given central cell.
We find that the combined structure is well described by only one parameter $\langle v \rangle$,
the mean number of extra bonds between adjacent layers.

\vskip 20pt
\noindent{\bf Acknowledgements:}
T. Aste acknowledges partial support from the European Union (TMR contract  ERBFMBICT950380).

\appendix

\section{Distribution of bond distances of cells all at the same shell distance}\label{ComDis}

Here, we estimate the quantity $N(j,t)$.  
First consider the layer of cells at a shell distance $j$ from the central
cell. 
Let  $P(t | j)$ be the conditional probability of finding, among all the cells at shell distance $j$  
from the central cell,  a cell at a bond distance $t$.  
This probability can be written  in terms of $N(j,t)$ as $P(t | j) = N(j,t)/ [ \sum_t N(j,t)]$.  
Let $w(i)$ be the probability to require $2+i$ bonds in the path between a layer and the next.
Therefore, any cell in layer $j$ is connected, in the bond tree, to a nearest cell the layer $j-1$ 
through a (shortest) path of $2+i$ bonds, with probability $w(i)$.  
Thus, 
\begin{equation}
P(t | j) = \sum_{i=0}^{t-2j+3} w(i) P( t-i-2 | j-1) \;\;\;\; (\mbox{for $j > 2$}) .
\label{Pjt}
\end{equation} 

Let us now calculate $P(t | 1)$ and $P(t |2)$. 
Cells sharing an edge with the central cell  ($j=1$) are at a bond distance zero. 
Thus, $P(t | 1) = \delta_{t,0})$. 
Cells of the second layer  ($j=2$) are connected to the nearest cell in the
first layer through a path of $1+i$ bonds with the same probability $w(i)$
($i=0$ labels those cells connected by one bond directly to the central cell). 
Thus,
\begin{equation} 
P(t | 2) = \sum_{i=0}^{t-1} w(i) P( t-i-1| 1) = w (t-1) \;\;\;\; .
\label{P2t}
\end{equation}
This fixes the initial condition in Eq.(\ref{Pjt}).

The weights $w(i)$ can be in principle any discrete probability distribution. 
The simplest case is when $w(i) =\delta_{i,0}$. 
In this case, Eq.(\ref{Pjt}) has the solution $P(t | j) = \delta_{t,2j-3}$.  
This solution has values
different from zero only for $t$ odd, it corresponds to a structure with minimal bond 
length throughout, i.e. an hexagonal tiling, except for the arbitrary central cell.

A realistic probability function for the weights is the Poisson
distribution of mean $\langle v \rangle$:
\begin{equation}
w(i) = {e^{-\langle v \rangle} \langle v \rangle^i \over i!  } \;\;\; .
\label{w}
\end{equation}

With this Ansatz, Eq.(\ref{Pjt}) has the solution
\begin{equation}
P(t | j) = {e^{-\langle v \rangle(j-1)} [\langle v \rangle (j-1)]^{t-2j+3} \over (t-2j+3)!  }\;\;\;\;\; .
\label{Pjt_sol}
\end{equation} 
The number of couples of cells at shell distance $j$ and bond distance $t$ is $N(j,t) = N K_j P(j |t)$.
As function of $t$ (with $j$ as a parameter) $N(j,t)$ is a shifted Poisson distribution
with average $\bar t = \langle v \rangle (j-1) +2j-3$ and  variance $\sigma^2 = \langle v \rangle (j-1)$).


\begin{thebibliography}{10}

\bibitem{dAT}
\newblock {D'A.W. Thompson, {\it On Growth and Form} Cambridge Univ. Press. (1917, 1942), ch.7.}

\bibitem{WR84}
\newblock {D. Weaire and N. Rivier, {\it Contemp. Physics} {\bf
25} (1984) 59.}

\bibitem{Stavans:0}
J. Stavans, {\it Rep. Prog. Mod. Phys.} {\bf 56} (1993) 733. 

\bibitem{Alexander}
\newblock {J. W. Alexander, {\it Ann. Math.} {\bf 31} (1930) 292.}


\bibitem{DR}
B. Dubertret and N. Rivier, {\it Biophys. Journal} {\bf 73} (1997) 38.

\bibitem{bd98}
\newblock {B. Dubertret, {\it Etudes th\'eoriques des syst\`emes cellulaires d\'esordonn\'es.
 Application \`a la mod\'elisation des tissus \'epith\'eliaux},
 PhD Thesis, (Universit\'e Louis Pasteur  Strasbourg, France 1998).}

\bibitem{dst97}
\newblock {B. Dubertret and K.Y. Szeto and W.Y. Tam, In preparation (1997)}


\bibitem{RivPc}
N. Rivier,  unpublished notes (1985); seminar, Imperial College (1986).


\bibitem{LeCearPc}
J. P. Troadec, unpublished notes (1985). 

\bibitem{Fortes}
M.A. Fortes and P.Pina, {\it Philos. Mag. B} {\bf 67}, 263 (1993). 

\bibitem{AsBoRi}
T. Aste, D. Boose and N. Rivier, {\it Phys. Rev. E} {\bf 53}, (1996) 6181.


 

\bibitem{AST}
T. Aste, K. Y. Szeto and W. Y. Tam, {\it Phys. Rev. E} {\bf 54}, (1996) 5482.

\bibitem{ShapeM:1}
T. Aste and N. Rivier, in {\it Shape Modelling and Applications} 
(IEEE Computer Society Press, 1997), p.2-9.


\bibitem{Aboav}
D. A. Aboav, {\it Metalloraphy} {\bf 3}, (1974) 383; {\bf 13}, (1980) 43.

\bibitem{Wea}
D. Weaire, {\it Metallography} {\bf 7}, (1974) 157.    

\bibitem{RiDuAsHo}
N. Rivier, B. Dubertret, T. Aste and H. M. Ohlenbusch, (1997) In preparation.

\bibitem{Del-LeCaer}
R. Delannay and G. LeCa\"er, {\it J. Physique I} {\bf 5} (1995) 1417.

\bibitem{AstRW97}
T. Aste, {\it Phys. Rev. E} {\bf 55}, (1997) 6233.

\end{thebibliography}
\end{document}